\documentclass[pre,twocolumn]{revtex4}
\usepackage{psfrag}
\usepackage{amssymb,amsmath,amsthm}
\usepackage[dvips]{graphicx}
\usepackage{verbatim}
\begin{document}
\title{ORDER VIA NONLINEARITY IN RANDOMLY CONFINED BOSE GASES}
\author{ROBERT GRAHAM and AXEL PELSTER}
\affiliation{Universit{\"a}t Duisburg-Essen, Campus Duisburg, Fachbereich Physik, 
Lotharstra{\ss}e 1, 47048 Duisburg, Germany\\
e-mail: {\tt robert.graham@uni-due.de, axel.pelster@uni-due.de}}      
\begin{abstract}
A  Hartree-Fock mean-field theory of a weakly interacting Bose-gas in a  quenched white noise disorder 
potential is presented. A direct continuous transition from the normal gas to a localized Bose-glass phase 
is found which has localized short-lived excitations with a gapless density of states and vanishing superfluid 
density. The critical temperature of this transition is as for an ideal gas undergoing Bose-Einstein condensation. 
Increasing the particle-number density a first-order transition from the localized state to a superfluid phase 
perturbed by disorder is found. At intermediate number densities both phases can coexist.

\bigskip

\noindent {\it Keywords:} Bose gas; quenched disorder; Hartree-Fock mean-field theory; phase diagram;
Bose-glass order parameter; Bose-glass phase.

\bigskip

\noindent {\it Dedicated to Hermann Haken on the occasion of his 80th birthday.}
\end{abstract}
\maketitle
\noindent {\bf 1. Introduction} \smallskip

\noindent After the successful realization of the Bose--Hubbard model in systems of ultracold atoms in perfectly 
periodic standing wave laser fields 
[Greiner {\it et al.}, 2002] 
the interest in the `dirty boson problem' 
[Fisher {\it et al.}, 1989; Giamarchi \& Schulz, 1987; Giamarchi \& Schulz, 1988]  
has also had a strong recent revival 
[Wang {\it et al.}, 2004]. 
In this case cold bosonic atoms in potentials with quenched 
disorder are considered. The disorder appears either naturally as, e.g., in magnetic wire traps
[Folman {\it et al.}, 2002; Schumm {\it et al.}, 2005; Fort\'agh \& Zimmermann, 2007] 
or it may be created artificially and controllably as, e.g., by the use of laser speckle fields 
[Dainty, 1975; Lye {\it et al.}, 2005; Cl\'ement {\it et al.}, 2005; Schulte {\it et al.}, 2005]. 
Historically, the dirty boson problem first arose in the context of superfluid Helium in Vycor 
[Wong {\it et al.}, 1990]. 
These classical experiments stimulated a number of papers in which the Bogoliubov 
mean-field theory of weakly interacting Bose gases was extended to superfluid Bose gases moving in a weak 
disorder-potential 
[Huang \& Meng, 1992; Giorgini {\it et al.}, 1994; Kobayashi \& Tsubota, 2002;
Lopatin \& Vinokur, 2002; Astrakharchik {\it et al.}, 2002; Falco {\it et al.}, 2007]. 
In these papers the perturbation and 
persistence of the superfluid phase in the presence of disorder was studied in detail. The problem of strong 
disorder potentials could not be dealt with in these theories, however, as this needs more 
sophisticated approaches  
[Hertz {\it et al.}, 1979; Navez {\it et al.}, 2007; Yukalov \& Graham, 2007; Yukalov {\it et al.}, 2007].
Apart from this work the contemporary problem of cold atoms in quenched random 
potentials has also been dealt with in numerical simulations of lattice systems 
[Lee \& Gunn, 1990; Ma {\it et al.}, 1993; Singh \& Rokhsar, 1994; 
Damski {\it et al.}, 2003; Sanpera {\it et al.} 2004, Krutitsky {\it et al.}, 2006]. 
An exception is the one-dimensional case of strongly repelling `dirty bosons' which 
can be solved exactly by a mapping, via fermionization, to the problem of the Anderson localization of 
non-interacting fermions 
[De Martino {\it et al.}, 2005; Krutitsky {\it et al.}, 2008].

\noindent
In the present paper we develop an analytical mean-field approach to the dirty boson problem of a weakly 
interacting Bose gas in a quenched random potential which is bounded from below, i.e.
we assume that all realizations  $V({\bf x})$ are larger than or equal to 
a fixed lower bound $V_0$. We shall work out our theory 
for the quite general model of a disorder potential 
$V({\bf x})$ where the ensemble average, in the following denoted by 
$\overline{\phantom{V}\hspace*{-1mm}\cdots\hspace*{-1mm}\phantom{V}}$,
vanishes, i.e.
$\overline{V({\bf x}_1)}= 0$, and, e.g., the second-order correlation function is given according to
$\overline{V({\bf x}_1) V({\bf x}_2)} = R^{(2)} ({\bf x}_1,{\bf x}_2)$. Our mean-field theory requires
no further assumptions about higher correlation functions $\overline{V({\bf x}_1) \cdot \ldots \cdot V({\bf x}_n)}$.
In the special case of a short-ranged correlation function 
$R^{(2)} ({\bf x}_1,{\bf x}_2)= R \,\delta ({\bf x}_1-{\bf x}_2)$
we find that, in addition to the superfluid phase studied in earlier work, for sufficiently 
strong disorder there exists a completely localized phase, either separately, at low number density, or in 
coexistence with a global condensate
at intermediate number densities. Our approach, from the start, differs qualitatively from the earlier 
mean-field approaches as we define and evaluate {\it two} basic order parameters of the theory
from disorder-averages over correlation functions of the underlying Bose fields $\psi^* ({\bf x}, \tau)$
and $\psi ({\bf x}, \tau)$. Here $\tau = i t$ stands for the imaginary time which is periodic with
period $\hbar \beta$, $\beta = 1/ k_B T$ denoting the reciprocal temperature.
All quantum mechanical expectation values here will be understood as being time-ordered with
respect to imaginary time $\tau$.
For characterizing the superfluid phase we use the usual condensate density $n_0$
which is defined by the spatial long-range limit of the 2-point function according to 
\begin{equation}
\lim_{|{\bf x}-{\bf x'}| \to \infty} \overline{\langle \psi ( {\bf x}, \tau ) \psi^* ( {\bf x'}, \tau^+ ) 
\rangle} = n_0 \, .
\end{equation}
Here and in the following the infinitesimally shifted imaginary time 
$\tau^+ = \tau + \eta$ with $\eta \downarrow 0$ is necessary to guarantee the normal ordering within
the underlying functional integral representation of the 2-point function.
But, in addition,  we also take into 
account the possible existence of a Bose-glass phase by introducing a corresponding separate 
order parameter $q$
which is similar to the Edwards-Anderson order parameter of a spin glass 
[Edwards \& Anderson, 1975]:
\begin{eqnarray}
\lim_{|{\bf x}-{\bf x'}| \to \infty}
\overline{\left|\langle  \psi ( {\bf x}, \tau ) \psi^* ( {\bf x'}, \tau^+ )  \rangle \right|^2 }
= (q+ n_0)^2 \, .
\end{eqnarray}
Note that the latter, at $T=0$, also follows from the temporal long-range correlation 
in direct analogy to the theory of quantum spin glasses 
[Read {\it et al.}, 1995; Sachdev, 1999]:
\begin{equation}
\lim_{|\tau- \tau'| \to \infty} \overline{\langle \psi ( {\bf x}, \tau ) \psi^* ( {\bf x}, \tau' ) \rangle} = q + n_0 \, .
\end{equation}
This procedure of introducing a separate Bose-glass order parameter $q$ is 
indispensable, we argue here, since in a sufficiently large system 
mutually disconnected local mini-condensates of bosons 
can form at sufficiently low temperature. Clearly, if such local, disconnected mini-condensates occur, with or 
without a surrounding sea of superfluid bosons, they cannot be described by the usual global mean field of a 
Bose--Einstein condensate, while they are well captured by an Edwards-Anderson-like order parameter. The definition 
and use of such a Bose-glass order parameter in the context of an otherwise very familiar model and the discussion 
of the simple results for the Bose-glass phase obtained in this way is the main goal of the present work.

\bigskip

\noindent {\bf 2. Replica Method} \smallskip

\noindent
Technically, our approach makes use of the replica method which has been established, over the last few decades, 
as a reliable tool for treating disordered systems 
[Mezard {\it et al.}, 1987; Fischer \& Hertz, 1991; Dotsenko, 2001]. 
We start with the functional integral for the grand-canonical partition function
\begin{eqnarray}
\label{ZP18}
{\cal Z} = \oint {\cal D} \psi^*  \oint {\cal D} \psi \, e^{-{\cal A}[\psi^*,\psi]} \, ,
\end{eqnarray}
where we employ units with $\hbar = 1$.
The integration is performed over all Bose fields $\psi^*({\bf x},\tau), \psi({\bf x},\tau)$ 
which are periodic in imaginary time $\tau$ with period $\beta$. The Euclidean action in standard notation is given  by
\begin{eqnarray}
\label{ZP19}
&& {\cal A} [\psi^*,\psi] = \int_0^{\beta} d \tau \int d^D x \,
\Big\{ \psi^* ({\bf x},\tau) \Big[ \frac{\partial}{\partial \tau}
- \frac{1}{2 M} {\bf \Delta} 
\nonumber \\ 
&&
+ V({\bf x})- \mu \Big] \psi ({\bf x},\tau) + \frac{g}{2}\,\psi^{*2} ({\bf x},\tau) \psi^2 ({\bf x},\tau) \Big\} \, ,
\end{eqnarray}
where $M$ denotes the particle mass, $\mu$ the chemical potential, and $g$ 
the strength of the contact interaction.
As the grand-canonical partition function ${\cal Z}$ still is a functional of the disorder potential
$V({\bf x})$, the corresponding thermodynamic potential follows from  the expectation value
with respect to disorder 
\begin{eqnarray}
\label{ZP20}
\Omega = - \frac{1}{\beta}\,\overline{\ln {\cal Z}} \, .
\end{eqnarray}
In general it is not possible to explicitly evaluate expression  (\ref{ZP20}),
as the averaging with respect to the disorder potential $V({\bf x})$ and the nonlinear
function of the logarithm do no commute:
$\overline{\ln {\cal Z}} \neq \ln  \overline{\cal Z}$.
An important method to perform the averaging procedure prescribed by (\ref{ZP20}) is provided by 
calculating the $N$th power of the grand-canonical partition function ${\cal Z}$ in the limit 
$N  \to 0$. Indeed, from 
${\cal Z}^N =  1 + N \, \ln {\cal Z} + \ldots$ 
the thermodynamic potential (\ref{ZP20}) is representable as
\begin{eqnarray}
\label{ZP23}
\Omega  = - \frac{1}{\beta}\,\lim_{N \to 0} \frac{\overline{{\cal Z}^N}-1}{N} \, .
\end{eqnarray}
The $N$-fold replication of the disordered Bose gas (\ref{ZP18}), (\ref{ZP19}) and a 
subsequent averaging with respect to the disorder potential $V({\bf x})$ can be worked out by
using the characteristic functional of the disorder potential $V({\bf x})$. Due to the above mentioned
assumptions about the statistical properties of the disorder potential $V({\bf x})$,
its characteristic functional is of the following form:
\begin{eqnarray}
\label{ZP12}
&& \overline{\exp \left\{ i \int d^D x \,j({\bf x})V({\bf x}) \right\}} = 
\exp \left\{ \sum_{n=2}^\infty \frac{i^n}{n!} \int d^D x_1 \cdots \right. \nonumber \\
&& \left.  \times \int d^D x_n \,R^{(n)} ({\bf x}_1,\ldots ,{\bf x}_n ) \,j({\bf x}_1) \cdots j({\bf x}_n) \right\} \, .
\end{eqnarray}
\begin{widetext}
The resulting $N$-fold replicated imaginary-time action of the 
weakly interacting Bose gas in a random potential reads
%
\begin{eqnarray}
&& {\cal A}^{(N)} [\psi^*,\psi]=
\int_0^{\beta} d \tau \int d^D x \sum_{\alpha=1}^N \,
\left\{ \psi^*_{\alpha} ({\bf x},\tau)\left[ \frac{\partial}{\partial \tau}
- \frac{1}{2 M} {\bf \Delta} - \mu \right] \psi_{\alpha} ({\bf x},\tau) 
+ \frac{g}{2}\,\left| \psi_{\alpha}({\bf x},\tau) \right|^4 \right\} 
+ \sum_{n=2}^\infty \frac{(-1)^{n-1}}{n!} 
\label{ZP29}
\\
&&\times
\int_0^{\beta} d \tau_1 \cdots \int_0^{\beta} d \tau_n
\int d^D x_1 \cdots\int d^D x_n  
\int d^D x_n \sum_{\alpha_1=1}^N \cdots \sum_{\alpha_n=1}^N
R^{(n)} ({\bf x}_1,\ldots ,{\bf x}_n ) \,
\left| \psi_{\alpha_1} ({\bf x}_1,\tau_1) \right|^2 \cdots 
\left| \psi_{\alpha_n} ({\bf x}_n,\tau_n) \right|^2 \, ,
\nonumber
\end{eqnarray}
where the indices $\alpha, \alpha'$ label the replicated copies of the Bose field.
Note that, by definition, all cumulant functions $R^{(n)} ({\bf x}_1,\ldots ,{\bf x}_n )$
are symmetric with respect to their arguments ${\bf x}_1,\ldots ,{\bf x}_n$.
Thus, in leading order $n=2$ the random potential leads  to a residual attractive interaction 
between the replica fields $\psi^*_{\alpha} ({\bf x},\tau)$, $\psi_{\alpha} ({\bf x},\tau)$ which 
is, in general, bilocal in both space and imaginary time. The replicated action
(\ref{ZP29}) is not only the starting point for calculating the disorder average of the grand-canonical 
potential  according to (\ref{ZP23}) and  
\begin{eqnarray}
\label{ZP24}
\overline{{\cal Z}^N} = \left\{ \prod_{\alpha=1}^N \oint {\cal D} \psi^*_{\alpha}  
\oint {\cal D} \psi_{\alpha} \right\}
\, e^{-{\cal A}^{(N)}[\psi^*,\psi]} \, ,
\end{eqnarray}
but also for the calculation of all disorder-averaged correlation functions of 
$\psi ( {\bf x}, \tau )$, $\psi^* ( {\bf x}, \tau )$. 
For instance, the 2-point function  can be represented within the replica formalism as 
\begin{eqnarray}
&&\overline{\langle \psi ( {\bf x}, \tau ) \psi^* ( {\bf x'}, \tau' ) \rangle} = 
\lim_{N \to 0} \frac{1}{N} \sum_{\alpha = 1}^N 
\left\{ \prod_{\alpha'=1}^N \oint {\cal D} \psi^*_{\alpha'}  \oint {\cal D} \psi_{\alpha'} \right\} 
\psi_{\alpha} ( {\bf x}, \tau ) \psi^*_{\alpha} ( {\bf x'}, \tau' ) \, e^{-{\cal A}^{(N)}[\psi^*,\psi]} \, .
\label{2P}
\end{eqnarray}

\bigskip

\noindent {\bf 3. Hartree-Fock Mean-Field Equations} \smallskip

\noindent
Now we shall analyze the action (\ref{ZP29}) 
in a Hartree-Fock (HF) approximation with respect to the direct and disorder-mediated interactions. To this end
we split the Bose-fields $\psi_{\alpha}$ into a background $\Psi_{\alpha}$ and fluctuations  $\delta\psi_{\alpha}$.
Due to their smallness we only keep terms up to the quartic order in the fluctuations  $\delta\psi_{\alpha}$,
which are then Gaussian factorized.
With this we obtain the so-called Gross-Pitaevskii equation for the background field, which is 
modified by the disorder-term, in the form
\begin{eqnarray}
\label{MF1}
&&
\left\{ \frac{\partial}{\partial \tau} - 
\frac{1}{2 M} {\bf \Delta} - \mu + g \,\Sigma_\alpha ({\bf x},\tau) +
g \langle \delta \psi_{\alpha}({\bf x},\tau) \, \delta \psi_{\alpha}^* ({\bf x},\tau^+) \rangle\right\} 
\Psi_{\alpha} ({\bf x},\tau) \\ \nonumber
&&=
\int_0^{\beta} d \tau'\int d^D x' \sum_{\alpha'=1}^N \,R^{(2)} ({\bf x}, {\bf x'})\Big\{ 
 \,Q_{\alpha \alpha'}  ( {\bf x}, \tau;  {\bf x'}, \tau')
\Psi_{\alpha'} ({\bf x'},\tau')
+ \langle \delta \psi_{\alpha'}({\bf x'},\tau') \, \delta \psi_{\alpha'}^* ({\bf x'},\tau'^+) \rangle
\Psi_{\alpha}({\bf x},\tau)  \Big\} + \ldots \, .
\end{eqnarray}
Here we introduced the mean-fields
\begin{eqnarray}
\label{MF4}
Q_{\alpha \alpha'} ( {\bf x}, \tau;  {\bf x'}, \tau') = \Psi_\alpha ({\bf x}, \tau) \,
\Psi_{\alpha'}^* ({\bf x'}, \tau') +
\langle \delta \psi_{\alpha}({\bf x},\tau) \, \delta \psi_{\alpha'}^* ({\bf x'},\tau') \rangle
\, , \hspace{0.5cm}
\Sigma_\alpha ({\bf x},\tau) & = & Q_{\alpha \alpha} ( {\bf x}, \tau;  {\bf x},\tau^+) \,,   
\end{eqnarray}
where the expectation values are calculated with respect to the 
effective quadratic action of the fluctuations 
\begin{eqnarray}
\label{EFF2}
&&\hspace*{-3mm}
\tilde{\cal A}^{(N,2)}[\delta \psi^*,\delta \psi] = \int_0^{\beta} d \tau \int d^D x \sum_{\alpha=1}^N \,\Bigg\{
\delta \psi^*_{\alpha} ({\bf x},\tau) \left[ \frac{\partial}{\partial \tau}
- \frac{1}{2 M} {\bf \Delta} - \mu+ 2 g\,\Sigma_\alpha ( {\bf x},\tau) \ \right] 
\delta \psi_{\alpha} ({\bf x},\tau) - \int_0^{\beta} d \tau \int_0^{\beta} d \tau'\\
&&\hspace*{-3mm}
\times  \int d^D x \int d^D x' \sum_{\alpha,\alpha'=1}^N R^{(2)} ({\bf x}, {\bf x'})
\Bigg\{ \Sigma_\alpha ( {\bf x},\tau) \,
\delta \psi^*_{\alpha'} ({\bf x'},\tau')\, \delta \psi_{\alpha'} ({\bf x'},\tau')
+ Q_{\alpha \alpha'} ( {\bf x}, \tau;  {\bf x'}, \tau') \, \delta \psi_{\alpha'} ({\bf x'},\tau') \, 
\delta \psi^*_{\alpha} ({\bf x},\tau) 
\Bigg\} + \ldots \, . \nonumber
\end{eqnarray}
\end{widetext}
Note that the higher-order terms in the ellipsis of Eqs.~(\ref{MF1}) and (\ref{EFF2})
stem from cumulant functions $R^{(n)}$ with $n \ge 3$. 
We observe that Eq.~(\ref{MF4}) defines 
$Q_{\alpha \alpha'}$ as a mean field related to disorder and
$\Sigma_{\alpha}$ essentially as the total particle number density $n_{\alpha}$ in the replica index $\alpha$, 
i.e.~$\Sigma_{\alpha} = gn_{\alpha}$. 
It should be noted that the HF approximation leading to (\ref{MF1}) and (\ref{EFF2}) is much simpler but also
more restrictive than the Hartree-Fock-Bogoliubov approximation in which anomalous correlations
$\delta \psi \delta \psi$ and $\delta \psi^* \delta \psi^*$ would also appear. However, not too far below
the transition temperature the influence of such terms is expected to be small 
[Goldman {\it et al.}, 1981; Huse \& Siggia, 1982].

\noindent
In the following we solve our HF approximation 
for the special case of a delta-correlated disorder potential,
i.e. we assume $R^{(2)} ({\bf x},{\bf x'})= R \,\delta ({\bf x} - {\bf x'})$ from now on. 
In that case
the mean-field equations have a replica symmetric solution
\begin{eqnarray}
\Psi_{\alpha} ({\bf x},\tau) = \sqrt{n_0} \, &,& \quad 
\Sigma_\alpha ({\bf x},\tau) = \Sigma \, ,  \nonumber \\ 
Q_{\alpha \alpha'} ( {\bf x}, \tau;  {\bf x'}, \tau')  &=& Q ( \tau - \tau' ) \, \delta_{\alpha \alpha'} + q  + n_0
\label{AN1}
\end{eqnarray}
with the Matsubara decomposition
\begin{equation}
Q ( \tau - \tau' ) = \frac{1}{ \beta}\,\sum_{m=-\infty}^{\infty} Q_m \,e^{-i \omega_m (\tau- \tau')} 
\label{AN2}
\end{equation}
and the bosonic Matsubara frequencies $\omega_m=2\pi m/ \beta$. With this ansatz 
cumulant functions $R^{(n)}$ with $n \ge 3$ in the
Eqs.~(\ref{MF1}) and (\ref{EFF2}) do not contribute in the replica limit $N \rightarrow 0$.
Furthermore, the expectation values 
in (\ref{MF1}) and (\ref{MF4}) are evaluated as follows.
The fluctuation action (\ref{EFF2}) is of the form
\begin{widetext}
\begin{eqnarray}
\hspace*{-0.2cm}
\tilde{\cal A}^{{(N,2)}} [\delta \psi^*,\delta \psi] = \int_0^{\hbar \beta} \hspace*{-1mm} d \tau \int_0^{\hbar \beta} 
\hspace*{-1mm}{d \tau'}\int d^D x\int d^D x'
\sum_{\alpha=1}^N \sum_{\alpha'=1}^N \frac{1}{2} \Big( \delta \psi_\alpha^* ( {\bf x}, \tau) , 
\delta \psi_\alpha ( {\bf x}, \tau) 
\Big) G_{\alpha \alpha'}^{-1} ( {\bf x}, \tau ; {\bf x'}, \tau')
\left( \begin{array}{@{}cc} \delta \psi_{\alpha'} ( {\bf x'}, \tau') \\
\delta \psi_{\alpha'}^* ( {\bf x'}, \tau') \end{array} \right) 
\end{eqnarray}
where the Fourier-Matsubara transform of the integral kernel 
$G_{\alpha \alpha'}^{-1} ( {\bf x}, \tau ; {\bf x'}, \tau')$ reads
\begin{eqnarray}
\label{DEKO4}
G_{\alpha \alpha'}^{-1} ( {\bf k}, \omega_m) = 
\left( \begin{array}{@{}cc} a( {\bf k}, \omega_m) & 0 \\ 0 & a^*( {\bf k}, \omega_m)
\end{array} \right)\, \delta_{\alpha \alpha'} +  \left( \begin{array}{@{}cc}  b( {\bf k}, \omega_m) & 0 \\ 0 &  
b^*( {\bf k}, \omega_m)
\end{array} \right)
\end{eqnarray}
with the quantities
\begin{eqnarray}
\label{AB}
a( {\bf k}, \omega_m) = - i \omega_m + \epsilon ( {\bf k}) - \mu + 2 g \Sigma - R Q_m   \, , \hspace*{1cm}
b( {\bf k}, \omega_m) = -  R ( q + \beta n_0   \, \delta_{m,0} )  \, ,
\end{eqnarray}
and the free dispersion $\epsilon ( {\bf k}) =  {\bf k}^2 / 2 M$.
Note that we have already performed in (\ref{DEKO4}) 
the replica limit $N \to 0$ which eliminates the Hartree contribution of the
disorder. The Fourier-Matsubara transform of the corresponding
Green function follows from algebraically inverting (\ref{DEKO4}) which yields
in replica limit $N \to 0$:
\begin{eqnarray}
G_{\alpha \alpha'} ( {\bf k}, \omega_m) =  \left( \begin{array}{@{}cc} {\displaystyle 
\frac{1}{a( {\bf k}, \omega_m)}} & 0 \\ 
0 & {\displaystyle \frac{1}{a^*( {\bf k}, \omega_m)}} \end{array} \right)\, \delta_{\alpha \alpha'}
- \left( \begin{array}{@{}cc} 
{\displaystyle \frac{b( {\bf k}, \omega_m)}{a( {\bf k}, \omega_m)^2}} & 0 \\ 
0 & {\displaystyle \frac{b^*( {\bf k}, \omega_m)}{a^*( {\bf k}, \omega_m)^2}} \end{array} \right) \, .
\end{eqnarray}
As the Green function contains expectation values according to
\begin{eqnarray}
G_{\alpha \alpha'} ( {\bf x}, \tau ; {\bf x'}, \tau')  = \left( 
\begin{array}{@{}cc} \langle \delta \psi_\alpha ( {\bf x}, \tau ) 
\delta \psi_{\alpha'}^* ( {\bf x'}, \tau' ) \rangle & 0 \\
0 & \langle \delta \psi_\alpha^* ( {\bf x}, \tau )\delta \psi_{\alpha'} ( {\bf x'}, \tau' ) \rangle 
\end{array} \right) \, ,
\end{eqnarray}
\end{widetext}
we arrive at the expression
\begin{eqnarray}
\label{EXP}
 \langle \delta \psi_\alpha ( {\bf x}, \tau ) \delta \psi_{\alpha'}^* ( {\bf x'}, \tau' ) \rangle 
&=& 
g_1  ( {\bf x}, \tau ; {\bf x'}, \tau' ) \,\delta_{\alpha \alpha'}
 \nonumber  \\
&&+ g_2  ( {\bf x}, \tau ; {\bf x'}, \tau' )\, ,
\end{eqnarray}
where the respective Fourier-Matsubara coefficients, in the replica limit $N \to 0$, read
\begin{eqnarray}
\label{G1}
\hspace*{-3mm}g_1  ( {\bf k}, \omega_m) &=&
\frac{1}{\beta 
\left[ - i  \omega_m + \epsilon({\bf k}) - \mu + 2\Sigma - R Q_m \right]} \, , \\ 
\label{G2}
\hspace*{-3mm}g_2  ( {\bf k}, \omega_m )&=& 
\frac{R  (q + n_0)\,\delta_{m,0}}{[\epsilon({\bf k}) - \mu + 2\Sigma - R Q_0 ]^2} \, .
\end{eqnarray}
Evaluating (\ref{MF4}) together with (\ref{AN1})--(\ref{G2}) determines the Matsubara coefficients
\begin{eqnarray}
\label{QM}
Q_m &=& - 2\pi R  \left( \frac{M}{2 \pi } \right)^{3} \\ \nonumber
&&- 2 \sqrt{\pi}
\left( \frac{M}{2 \pi} \right)^{3/2}
\sqrt{- i  \omega_m - \mu + 2\Sigma + \pi R^2 \left(\frac{M}{2 \pi} \right)^{3}} 
\end{eqnarray}
and the Bose-glass order parameter
\begin{eqnarray}
q & = & 
\frac{ \sqrt{\pi} R (q + n_0)}{\sqrt{- \mu + 2\Sigma - R Q_0}}\, \left( \frac{M}{2 \pi } \right)^{3/2} \, .
\label{q}
\end{eqnarray}
Inserting these results in  (\ref{MF1}) and (\ref{MF4}) and taking the replica limit $N\to 0$ we find the 
Gross-Pitaevskii equation
\begin{eqnarray}
\label{PMM}
( \mu - 2\Sigma + gn_0 + R Q_0 ) \, \sqrt{n_0} =  0
\end{eqnarray}
and the particle density
\begin{eqnarray}
\label{MAT1}
n = \frac{1}{g} \, \Sigma = q + n_0
+ \lim_{\eta\downarrow 0} \sum_{m = -\infty}^{\infty}\frac{Q_m}{\beta}
e^{i\omega_m\eta}\, .
\end{eqnarray}
The remaining Matsubara sum in (\ref{MAT1}) is evaluated together with (\ref{QM}) by using the zeta function
regularization method [Kleinert, 2004]. Applying both the Schwinger trick
\begin{eqnarray}
\label{SCH}
\frac{1}{a^\nu} =\frac{1}{\Gamma(\nu)}\int_0^{\infty} d s \,s^{\nu-1} \, e^{- a s} 
\end{eqnarray}
with the Gamma function $\Gamma$ and the Poisson sum formula
\begin{eqnarray}
\label{PF}
\sum_{m=-\infty}^\infty \delta (x-m) = \sum_{n=-\infty}^\infty e^{- 2 \pi i n x} 
\end{eqnarray}
yields the result
\begin{eqnarray}
\label{SUM}
\sum_{m=-\infty}^\infty \left( - i \hbar \omega_m + a \right)^\nu = 
\frac{\zeta_{\nu+1} \left( e^{- a \beta} \right)}{\beta^\nu \,\Gamma(- \nu)} 
\end{eqnarray}
with the standard polylogarithmic function
\begin{eqnarray}
\zeta_\nu (z) = \sum_{n=1}^\infty \frac{z^n}{n^\nu} \, .
\end{eqnarray}
With this we obtain finally the equation of state of the system
\begin{eqnarray}
\label{SI}
n = \frac{1}{g} \, \Sigma = q + n_0 + \left( \frac{M}{2 \pi  \beta} \right)^{3/2} \, \zeta_{3/2} \left(
e^{\beta \mu_r} \right) \, 
\end{eqnarray}
with the abbreviation $\mu_r(\mu) = \mu - 2gn(\mu) - \pi R^2  (M / 2\pi )^{3}$ for the renormalized 
chemical potential.  
Thus, Eqs.~(\ref{q}), (\ref{PMM}), (\ref{SI}), and (\ref{QM}) with $m=0$ form a closed set of nonlinear
equations which determine the types of order the bosons can establish in the underlying random potential
via their quantum statistics and their interaction.
\begin{figure}[t]
\includegraphics[scale=0.4]{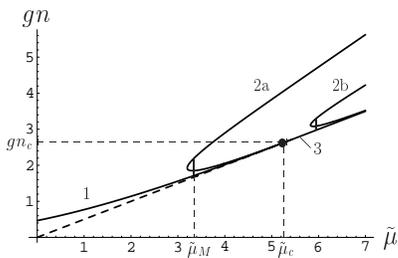}
\caption{
\label{ES}
Isotherm in the $(\tilde{\mu},gn)$-plane with 
$\tilde{\mu} = \mu -\pi R^2 (M / 2\pi)^{3}$ for two values of the disorder parameter $R$. 1: normal 
Bose gas; 2a, 2b: superfluid with the dimensionless disorder strength 
$\sqrt{\pi} R (M\beta^{1/3}/2\pi)^{3/2}$
set to 0.2 and 0.48, respectively; 3: Bose glass. The dimensionless coupling constant $g/(2\pi\beta^{1/3}/M)^{3/2}$ 
is fixed to 1.}
\end{figure}
\bigskip

\noindent {\bf 4. Phase Diagram} \smallskip

\noindent Let us first consider briefly the superfluid phase defined by $n_0 \neq 0$, 
as it is well-known from the earlier work 
[Huang \& Meng, 1992; Giorgini {\it et al.}, 1994; Kobayashi \& Tsubota, 2002;
Lopatin \& Vinokur, 2002; Astrakharchik {\it et al.}, 2002; Falco {\it et al.}, 2007]. 
To this end we solve (\ref{PMM}) for the condensate density  and use (\ref{QM}) for $m=0$, yielding
\begin{eqnarray}
n_0 = \frac{1}{g}\,\left[ \sqrt{-\mu_r}+ \sqrt{\pi} R \left( \frac{M}{2\pi} \right)^{3/2} \right]^2 \, .
\end{eqnarray}
Then we determine the Bose-glass order parameter from (\ref{q}):
\begin{eqnarray}
q = \frac{\sqrt{\pi} R (M/2\pi)^{3/2}}{g \sqrt{-\mu_r}}
\,\left[ \sqrt{-\mu_r}+ \sqrt{\pi} R \left( \frac{M}{2\pi} \right)^{3/2} \right]^2 \, .
\end{eqnarray}
Inserting all results in (\ref{SI}) 
we find the equation of state for fixed temperature:
\begin{eqnarray}
n&=& \frac{1}{g\sqrt{-\mu_r}} \, \left[\sqrt{-\mu_r}+ \sqrt{\pi} R \left( \frac{M}{2\pi} \right)^{3/2} 
\right]^3 \nonumber\\
&& +\left( \frac{M}{2 \pi  \beta} \right)^{3/2} \, \zeta_{3/2} \left(
e^{\beta \mu_r} \right) \, .
\end{eqnarray}
A  resulting isotherm is given in Fig.~\ref{ES}, where we plot $g n$ for two different disorder-strengths as a 
function of the chemical potential 
$\tilde{\mu} = \mu_r + 2 g n(\mu_r)=
\mu -\pi R^2 (M / 2\pi)^{3}$ both in units of $k_B T$.  
The isothermal compressibility $\kappa = n^{-2} \partial n/\partial \mu$ can be read off this graph. 
The part of the isotherm with $n_0 \neq 0$
consists of two branches and only exists above temperature- and 
disorder-dependent minimal values 
$\tilde{\mu} = \tilde{\mu}_m$ and $n = n_m$. The upper branch is locally stable, the lower branch is locally 
unstable. The stable branch starts at $\tilde{\mu} = \tilde{\mu}_m$ with infinite compressibility and, for 
large number density smoothly approaches from below the curve
$n=[\tilde{\mu} -3  \sqrt{\pi} R (M/2\pi)^{3/2}
\sqrt{\tilde{\mu}}+6\pi R^2 (M / 2\pi)^{3}] / g$. The lower unstable branch likewise starts  at 
$\tilde{\mu} = \tilde{\mu}_m$ with negative infinite compressibility, passes through a state of vanishing 
compressibility
and, for large $\tilde{\mu}$, approaches from above the line $n=\tilde{\mu} / 2g$. 
All this is qualitatively the same as in the case without disorder, where it is well-known that, 
in the HF-approximation, BEC appears via a first-order transition  
(cf.~[Flie{\ss}er {\it et al.}, 2001] and references therein). 
The latent heat of 
the transition comes from the additional repulsive Fock exchange interaction in the gas and the glass phase, 
which is absent for particles in the condensate. The new additional aspect here is that the corresponding
mean-field HF equations in the presence of disorder predict that $|\psi |^2 \neq 0$ implies also $q \neq 0$,
i.e. that the presence of a global condensate implies also the presence of mini-condensates if the
potential has a disordered component. The inverse is not true, however, as we shall now see. 

\noindent
To this end we turn to the branch of the isotherm with $n_0=0$. It consists of the normal Bose-gas branch
obtained by solving (\ref{PMM}) with $n_0=0$, then (\ref{q}) 
with $q=0$, and finding the equation of state from (\ref{SI}):
\begin{eqnarray}
n=\left( \frac{M}{2\pi\beta} \right)^{3/2}\zeta_{3/2}(e^{\beta\mu_r}) \, .
\end{eqnarray}
This part of the isotherm ends in a critical point at the 
critical particle number density $n_c= ( M/2 \pi \beta_c )^{3/2} \zeta (3/2)$ 
and chemical potential $\tilde{\mu}_c = 2g n_c $ or, equivalently, $\mu_r = 0$.
In addition, however, 
we find a Bose-glass part of the isotherm ($n_0=0, q \neq 0$) along
$n=\tilde{\mu}/ 2g$, i.e.  $\mu_r = 0$,
for $\tilde{\mu} > \tilde{\mu}_c$, $n > n_c$, which starts in the critical point with a 
continuous phase transition. It turns out to have the same critical temperature 
$T_c^0=2\pi [n/\zeta(3/2)]^{2/3}/ k_B M$
as the BEC transition of the ideal Bose gas, where $\zeta(\nu)=\zeta_{\nu}(1)$ denotes the Riemann
zeta function. This means that  the interaction has a negligible effect
at the formation of the independent mini-condensates, which
make up the Bose-glass phase.
Furthermore, it has the finite compressibility $\kappa = 1/2g n^2$ and 
is locally stable. This part of the isotherm is found by again solving (\ref{PMM}) for $n_0 = 0$, but now 
solving (\ref{q}) for arbitrary $q\neq 0$ by putting 
$\sqrt{\pi} R (M / 2\pi)^{3/2}= \sqrt{-\mu + 2\Sigma - R Q_0}$, which corresponds to $\mu_r =0$, so
the equation of state reads
\begin{eqnarray}
n = \frac{1}{2 g} \,\left[ \mu - \pi R^2 \left( \frac{M}{2 \pi} \right)^3\right] \, ,
\end{eqnarray}
and finally solving (\ref{SI}) by 
$q  = n-n_c$. The isotherm with the normal part and the Bose-glass part is also indicated in Fig.~\ref{ES}. 
Also indicated there is the value $\tilde{\mu}_M$ of $\tilde{\mu}$ for the coexistence between
superfluid and normal or Bose-glass, respectively, which is fixed by the 
Maxwell-construction. To the left (resp.\,right) of this value the normal or Bose-glass state 
(resp. superfluid state) are absolutely stable. Thus, for the two cases of the disorder strength
exhibited in Fig.~\ref{ES} only the case with the larger disorder permits a stable Bose-glass state.

\noindent
We shall now focus our discussion on properties of the Bose-glass phase. As we saw it is characterized by the 
absence of off-diagonal long-range order in space ($n_0=0$) but the presence of long-range order in time 
($q\neq 0$), like in a spin-glass phase. Note that the appearance of $q$ spontaneously breaks the continuous 
$U(1)^{N}$-symmetry for $N$-replicas to a joint $U(1)$-symmetry. The latter is only spontaneously broken if also 
$n_0 \neq 0$.

\noindent
To see that the superfluid density vanishes in the Bose-glass phase we apply an infinitesimal Galilean boost with 
velocity $\delta{\bf u}$, using the Galilei-invariance of the distribution of the random potential $V({\bf x})$, 
and determine the linear response of the momentum density 
$\overline{\delta{\bf g}}= M n_n\delta{\bf u}$, which defines the normal gas density $n_n$. A direct evaluation 
yields $n_n = g_1 ({\bf x},\tau; {\bf x},\tau^+) 
+ g_2 ({\bf x},\tau; {\bf x}, \tau^+)$,
thus the normal density contains a Bose-glass contribution via $g_2$. The superfluid density $n_{s}$,
which is defined by $n_{s} = n- n_n$, turns out to be 
equal to $n_0$ which, as we have seen, vanishes in the pure localized phase. 

\noindent
The localization of the Bose-glass states can be inferred from the spatial exponential fall-off of the correlation 
function $g_2({\bf x},\tau ; {\bf x'}, \tau')$ describing correlations of the locally
condensed component. Eq.~(\ref{G2}) 
allows us to extract the temperature-independent localization length
$\xi_x = 2\pi/ M^2 R$.
Since the HF-approximation is an effective free-particle theory, this localization length is independent of both 
the number density $n$ and the interaction strength $g$. The same localization length is found by studying  the 
Anderson localization of a particle of mass $M$ in a spatial white-noise potential 
[Edwards \& Muthukumar, 1988]. 

\noindent
From (\ref{G1}) and (\ref{QM}) follows the spectral function 
$A_1(\omega, {\bf k})= -2 \Im [g_1 (\omega_m, {\bf k}) |_{i\omega_m = \omega - i 0}]$ of the excitations 
\begin{equation}
A_1(\omega, {\bf k})
=
\frac{2\pi R\overline{\rho(\omega)}}
{\left(\omega +\mu_r - \frac{{\bf k}^2-\xi_x^{-2}}{2M}\right)^2 
+\left(\frac{{\bf k}}{M\xi_x}\right)^2}
\label{A}
\end{equation}
with the density of states 
\begin{eqnarray}
\overline{\rho (\omega)}
&=& \frac{2}{\sqrt{\pi}}\left(
\frac{M}{2\pi}\right)^{3/2}                                
\sqrt{\omega +\mu_r} \,\Theta (\omega +\mu_r) \, .
\label{eq:24}
\end{eqnarray}
Thus, the excitations are resonances with energies larger than $-\mu_r = - \mu + 2gn + 1/ 2 M \xi_x^2$, 
width $\gamma = |{\bf k}|/ M\xi_x$, and wave numbers larger than the inverse localization length $\xi_x$. 
The size $-\mu_r$ of the energy gap in the normal phase and the value of the chemical potential 
$\mu=2gn + 1 / 2 M \xi_x^2$ in the gapless localized phase contain the mean-field shift due to the repulsive 
interaction but also the quantum mechanical localization energy of the particles of mass $M$.
The gap $-\mu_r$ in the normal phase disappears at the transition to the Bose-glass phase and remains zero 
in that phase. For $\omega\to 0$ the density of states then goes to zero as $\sqrt{\omega}$.
Note that our finding differs qualitatively from the lattice case where the density of states remains non-zero
in the limit $\omega \rightarrow 0$ [Fisher {\it et al.}, 1989;  Ma {\it et al.}, 1993; Singh \& Rokhsar, 1994].
We expect that a thorough study of the continuum limit of the lattice system would show that this density
of states tends to zero in that limit and, thereby, permit to confirm our result.
\begin{figure}[t]
\includegraphics[scale=0.4]{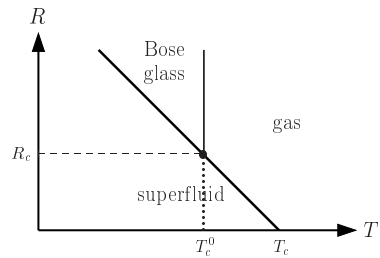}
\caption{
\label{rtdia}
Schematic phase diagram in the disorder-temperature plane where density and s-wave scattering length are kept 
fixed. Thick and thin lines represent first order and continuous phase transitions,
respectively.}
\end{figure}

\bigskip

\noindent {\bf 5. Conclusion} \smallskip

\noindent In conclusion we have described a simple mean-field HF approach to the weakly interacting disordered Bose-gas. 
Within a single approach this leads to the existence and possible coexistence of a superfluid phase and a 
localized Bose-glass phase. To illustrate our results further, let us briefly discuss the system also in the 
temperature-disorder $(T,R)$-plane at fixed particle-number density (see Fig.~\ref{rtdia}). At $R=0$ there is 
a critical temperature $T_c$ at which the superfluid (stable for $T<T_c$) and the normal Bose-gas 
(stable for $T>T_c$) coexist. Note that, due to the weak repulsive interaction with  $a n^{1/3}\ll 1$ 
(where $a=Mg/4\pi$ is the s-wave scattering length), this $T_c$ is larger than the critical temperature of 
the ideal Bose gas $T_c^0$ by about $\Delta T_c = T_c - T_c^0 \approx 1.3\, a n^{1/3} T_c^0$ 
(cf.~[Kastening, 2004] 
and references therein). Increasing $R$ the isotherm in Fig.~\ref{ES} shifts in such a way (to the right) that 
$T_c(R)$ for fixed $n$ decreases. There is then a critical minimal value $R_c$ of $R$ at which 
$\tilde{\mu}_M = \tilde{\mu}_{c}$ above which $T_c(R)$ drops below $T_c^0$, which is the critical temperature for the 
appearance of the localized Bose-glass phase. For stronger disorder the normal high-temperature phase upon cooling 
first makes a continuous transition to the localized phase, which is then followed by a first-order transition 
to the Bose-condensed phase. These predictions should be amply testable in current state-of-the-art experiments.

\bigskip
\noindent{\bf Acknowledgments} \smallskip

\noindent This work has been supported by the German Research Foundation (DFG) within
the Collaborative Research Center SFB/TR\,12 {\it Symmetries and Universality in Mesoscopic Systems}.
\bigskip

\noindent {\bf References} \smallskip

\noindent 
Greiner, M., Mandel, O., Esslinger, T., H{\"a}nsch, T.W. \& Bloch, I. [2002] 
``Quantum Phase Transition from a Superfluid to a Mott Insulator in a Gas of Ultracold Atoms,"
{\it Nature} {\bf 415}, 39-44.

\noindent 
Fisher, M.P.A., Weichman, P.B., Grinstein, G. \& Fisher, D.S. [1989]
``Boson Localization and the Superfluid-Insulator Transition,"
{\it Phys. Rev.} {\bf B40}, 546-570.

\noindent
Giamarchi, T. \& Schulz, H.J. [1987]
``Localization and Interaction in One-Dimensional Quantum Fluids,"
{\it Europhys. Lett.} {\bf 3}, 1287-1293.

\noindent
Giamarchi, T. \& Schulz, H.J. [1988]
``Anderson Localization and Interaction in One-Dimensional Metals,"
{\it Phys. Rev.} {\bf B37}, 325-340.

\noindent
Wang, D.-W., Lukin, M.D. \& Demler, E. [2004]
``Disordered Bose-Einstein Condensates in Quasi-One-Dimensional Magnetic Microtraps,"
{\it Phys. Rev. Lett.} {\bf 92}, 076802-1-076802-4.

\noindent
Folman, R., Kr{\"u}ger, P., Schmiedmayer, J., Denschlag, J. \& Henkel, C. [2002]
``Microscopic Atom Optics: From Wires to an Atom Chip," 
{\it Adv. At. Mol. Opt. Phys.} {\bf 48}, 263-356.

\noindent
Schumm, T., Esteve, J., Figl, C., Trebbia, J.B., Aussibal, C., 
Nguyen, H., Mailly, D., Bouchoule, I., Westbrook, C.I. \& Aspect, A. [2005]
``Atom Chips in the Real World: the Effects of Wire Corrugation,"
{\it Eur. Phys. J.} {\bf D32}, 171-180.

\noindent
Fort\'agh, J. \& Zimmermann, C. [2007]
``Magnetic Micro Traps for Ultracold Atoms,"
{\it Rev. Mod. Phys.} {\bf 79}, 235-289 (2007).

\noindent
Dainty, J.C. (Editor) [1975]
``Laser Speckle and Related Phenomena" (Springer, Berlin).

\noindent
Lye, J.E., Fallani, L., Modugno, M.,  Wiersma, D.S.,  Fort, C. \& Inguscio, M. [2005]
``Bose-Einstein Condensate in a Random Potential,"
{\it Phys. Rev. Lett.} {\bf 95}, 070401-1-070401-4.

\noindent
Cl{\'e}ment, D., Var{\'o}n, A.F.,  Hugbart, M.,  Retter, J.A., Bouyer, P., Sanchez-Palencia, L.,
Gangardt, D.M., Shlyapnikov,  G.V. \&  Aspect, A. [2005]
``Suppression of Transport of an Interacting Elongated Bose-Einstein Condensate in a Random Potential,"
{\it Phys. Rev. Lett.} {\bf 95}, 170409-1-170409-4. 

\noindent
Schulte, T., Drenkelforth, S., Kruse, J.,  Ertmer, W., Arlt, J., Sacha, K.,  
Zakrzewski, J. \& Lewenstein, M. [2005]
``Routes Towards Anderson-Like Localization of Bose-Einstein Condensates in Disordered Optical Lattices,"
{\it Phys. Rev. Lett.}  {\bf 95}, 170411-1-170411-4.

\noindent
Wong, G.K.S., Crowell, P.A.,  Cho, H.A. \& Reppy, J.D. [1990]
``Superfluid Critical-Behavior in He-4-Filled Porous-Media," 
{\it Phys. Rev. Lett.} {\bf 65}, 2410-2413.

\noindent
Huang, K. \& Meng, H.-F. [1992]
``Hard-Sphere Bose-Gas in Random External Potentials,"   
{\it Phys. Rev. Lett.} {\bf 69}, 644-647 (1992).

\noindent
Giorgini, S., Pitaevskii, L. \& Stringari, S. [1994]
``Effects of Disorder in a Dilute Bose-Gas,"
{\it Phys. Rev.} {\bf B49}, 12938-12944.

\noindent
Kobayashi M. \& Tsubota, M. [2002]
``Bose-Einstein Condensation and Superfluidity of a Dilute Bose Gas in Random Potential,"
{\it Phys. Rev.} {\bf B66}, 174516-1-174516-7.

\noindent
Lopatin, A.V. \& Vinokur, V.M. [2002]
``Thermodynamics of the Superfluid Dilute Bose Gas with Disorder," 
{\it Phys. Rev. Lett.} {\bf 88}, 235503-1-235503-4.

\noindent
Astrakharchik, G.E.,  Boronat, J., Casulleras, J. \& Giorgini, S. [2002]
``Superfluidity Versus Bose-Einstein Condensation in a Bose Gas with Disorder," 
{\it Phys. Rev.} {\bf A66}, 023603-1-023603-4.

\noindent
Falco, G.M., Pelster, A. \& Graham, R. [2007]
``Thermodynamics of a Bose-Einstein Condensate with Weak Disorder,"
{\it Phys. Rev.} {\bf A75}, 063619-1-063619-11.

\noindent
Hertz, J.A., Fleishman,  L. \&  Anderson, P.W. [1979]
``Marginal Fluctuations in a Bose Glass,"
{\it Phys. Rev. Lett.} {\bf 43}, 942-946.

\noindent
Navez, P., Pelster, A. \& Graham, R. [2007]
``Bose Condensed Gas in Strong Disorder Potential With Arbitrary Correlation Length,"
{\it Appl. Phys.} {\bf B86}, 395-398.

\noindent
Yukalov, V.I. \& Graham, R. [2007]
``Bose-Einstein-Condensed Systems in Random Potentials,"
{\it Phys. Rev.}, {\bf A75}, 023619-1-023619-16.

\noindent
Yukalov, V.I.,  Yukalova, E.P.,  Krutitsky, K.V. \& Graham R. [2007]
``Bose-Einstein Condensed Gases in Arbitrarily Strong Random Potentials,"
{\it Phys. Rev.} {\bf A76}, 053623-1-053623-11.

\noindent
Lee, D.K.K \& Gunn, J.M.F. [1990]
``Bosons in a Random Potential: Condensation and Screening in a Dense Limit,"
{\it J. Phys.} {\bf C2}, 7753-7768.

\noindent
Ma, M., Nisamaneephong, P. \& Zhang, L. [1993]
``Ground State and Excitations of Disordered Boson Systems,"
{\it J. Low Temp.} {\bf 93}, 957-969.

\noindent
Singh, K.G. \& Rokhsar, D.S. [1994]
``Disordered Bosons: Condensate and Excitations,"
{\it Phys. Rev.} {\bf B49}, 9013-9023.

\noindent
Damski, B., Zakrzewski, J., Santos, L., Zoller, P. \& Lewenstein, M. [2003]
``Atomic Bose and Anderson Glasses in Optical Lattices," 
{\it Phys. Rev. Lett.} {\bf 91}, 080403-1-080403-4.

\noindent
Sanpera, A., Kantian, A., Sanchez-Palencia, L., Zakrzewski, J. \& Lewenstein, M. [2004]
``Atomic Fermi-Bose Mixtures in Inhomogeneous and Random Lattices: 
From Fermi Glass to Quantum Spin Glass and Quantum Percolation," 
{\it Phys. Rev. Lett.} {\bf 93}, 040401-1-040401-4.

\noindent
Krutitsky, K.V., Pelster, A. \& Graham, R. [2006]
``Mean-Field Phase Diagram of Disordered Bosons in a Lattice at Non-Zero Temperature,"
New J. Phys. {\bf 8}, 187-1-187-14.

\noindent
De Martino, A., Thorwart, M., Egger, R. \& Graham, R.
``Exact Results for One-Dimensional Disordered Bosons with Strong Repulsion," 
{\it Phys. Rev. Lett.} {\bf 94}, 060402-1-060402-4.

\noindent
Krutitsky, K.V., Thorwart, M., Egger, R. \& Graham R. [2008]
``Ultracold Bosons in Lattices with Binary Disorder,"
{\it Phys. Rev.} {\bf A} (in press); eprint: {\tt arXiv:0801.2343}

\noindent
Edwards, S.F. \& Anderson,  P.W. [1975]
``Theory of Spin Glasses,"
{\it J. Phys.}  {\bf F5}, 965-974.

\noindent
Read, N., Sachdev, S. \& Ye, J. [1995]
``Landau Theory of Quantum Spin-Glasses of Rotors and Ising Spins,"
{\it Phys. Rev.} {\bf B52}, 384-410.

\noindent
Sachdev, S. [1999]
{\it Quantum Phase Transitions}
(Cambridge University Press, Cambridge).

\noindent
Mezard, M., Parisi, G. \& Virasoro, M.A. [1987]
{\it Spin Glass Theory and Beyond}
(World Scientif\/ic, Singapore). 

\noindent
Fischer, K.H. \& Hertz, J.A. [1991] 
{\it Spin Glasses}
(Cambridge University Press, Cambridge).

\noindent
Dotsenko, V. [2001]
{\it Introduction to the Replica Theory of Disordered Statistical Systems}
(Cambridge University Press, Cambridge).

\noindent
Goldman, V.V., Silvera, I.F. \& Leggett, A.J. [1981]
``Atomic Hydrogen in an Inhomogeneous Magnetic Field: Density Profile and Bose-Einstein Condensation,"
{\it Phys. Rev.} {\bf B24}, 2870-2873.

\noindent
Huse, D.A. \& Siggia, E.D. [1982]
``The Density Distribution of a Weakly Interacting Bose Gas in an External Potential,"
{\it J. Low Temp.} {\bf 46}, 137- 149.

\noindent
Kleinert, H. [2004]
{\it Path Integrals in Quantum Mechanics, Statistics, Polymer Physics, and Financial Markets,} Fourth Edition
(World Scientific, Singapore).

\noindent
Flie{\ss}er, M., Reidl, J., Sz{\'e}pfalusy, P. \& Graham, R. [2001]
``Conserving and Gapless Model of the Weakly Interacting Bose Gas," 
{\it Phys. Rev.} {\bf A64}, 013609-1-013609-14.

\noindent
Edwards, S.F. \& Muthukumar, M. [1988]
``The Size of a Polymer in Random-Media,"
{\it J. Chem. Phys.} {\bf 89}, 2435-2441.

\noindent
Kastening, B. [2004]
``Bose-Einstein Condensation Temperature of a Homogenous Weakly 
Interacting Bose Gas in Variational Perturbation Theory Through Seven Loops,"
{\it Phys. Rev.} {\bf A69}, 043613-1-043613-8.

\end{document}